\documentclass[final,english]{bullsrsl}[2022/06/15]

\usepackage[latin1]{inputenc}
\usepackage[T1]{fontenc}

\usepackage{natbib} 
\usepackage{graphicx}

\hypersetup{
        colorlinks   = true, 
        urlcolor     = magenta, 
        linkcolor    = blue, 
        citecolor   = blue 
}

\newcommand\asts{\textit{AstroSat}}

\newcommand{\abd}{AB~Dor}
\newcommand{\cce}{CC~Eri}

\newcommand\arcm{\hbox{$^\prime$}}            
\newcommand\cts{count~s$^{-1}$}

\begin{document}
\title{\textit{AstroSat} investigation of X-ray flares on two active K-M systems: CC~Eri and AB~Dor}

\author[affil={1}, corresponding]{Subhajeet}{Karmakar}
\author[affil={2}]{Jeewan C.}{Pandey}
\author[affil={2}]{Nikita}{Rawat}
\author[affil={2}]{Gurpreet}{Singh}
\author[affil={1,3}]{Riddhi}{Shedge}

\affiliation[1]{Monterey Institute for Research in Astronomy (MIRA), 200 Eighth Street, Marina, California 93933, USA}
\affiliation[2]{Aryabhatta Research Institute of Observational Sciences (ARIES), Manora Peak, Nainital 263002, India}
\affiliation[3]{Monta Vista High School, 21840 McClellan Rd, Cupertino, CA 95014, USA}
\correspondance{subhajeet09@gmail.com, sk@mira.org}
\date{13th October 2020}
\maketitle

\begin{abstract}
  We present an X-ray and UV investigation of five X-ray flares detected on two active systems, \cce\ and \abd, using the \asts\ observatory.  The peak X-ray luminosities of the flares in the 0.3--7.0~keV band are found to be within 10$^{31-33}$~erg~s$^{-1}$. Preliminary spectral analysis indicates the presence of three and four-temperature corona for CC Eri and AB Dor, respectively, where the highest temperature is found to vary with flare. The flare temperatures peaked at 51--59 MK for \cce\ and 29--44~MK for \abd. The peak emission measures of the flaring loops are estimated to be $\sim$10$^{54}$ for \cce\ and $\sim$10$^{55}$~cm$^{-3}$ for \abd. Global metallic abundances were also found to increase during flares.
\end{abstract}


\section{Introduction}
\label{sec:intro}
Solar-type stars with a convective envelope above a radiative interior show a high level of magnetic activity \citep[][]{Patel-16-MNRAS-6, KarmakarS-18-IAUS-1, KarmakarS-19-BSRSL-2, Savanov-18-AstBu-13}. Binaries and multiple systems are even more magnetically active due to the large influence of the rotation of the individual stars by tidal interactions. Understanding the magnetic activities are very important as they provide useful information about the stellar dynamo theory and stellar spaceweather \citep[][]{Maehara-12-Natur-2, KarmakarS-21-MIRAN-6}.

Stellar flares are extreme dynamic behavior of the stellar atmosphere and are an important manifestation of magnetic activities \citep[][]{Pandey-15-AJ-6, Savanov-18-ARep-8, KarmakarS-18-BSRSL-2, KarmakarS-16-MNRAS-2, KarmakarS-22-AAS-3, KarmakarS-23-MNRAS-2}. In this paper, we investigated two highly active systems \cce\ and \abd\ observed with \asts\ satellite. \cce\ is a K7.5+M3.5 binary located at a distance of $\sim$11.5 pc, whereas \abd\ is known to be a K0+M8+M5+M5-6 quadrupole system located at $\sim$14.9 pc \citep[][]{Bailer-Jones-18-AJ-6}. Both the objects showed variations in X-ray and UV, and multiple flaring activities have been observed \citep[][]{Crespo-Chacon-07-A+A, KarmakarS-17-ApJ, KarmakarS-19-PhDT-2, DidelS-22-csss}. Using the simultaneous observations in UV and X-ray, our objective is to investigate the coronal and chromospheric features of the systems. 

We structured the paper as follows: Section~\ref{sec:redn} describes the observations and data reduction. Analysis and results from X-ray timing and spectral analysis are presented in Section~\ref{sec:result}. Finally, in Section~\ref{sec:concl}, we discuss the result and present conclusions.

\begin{figure*}
\centering
\includegraphics[height=6.2cm, angle=-0]{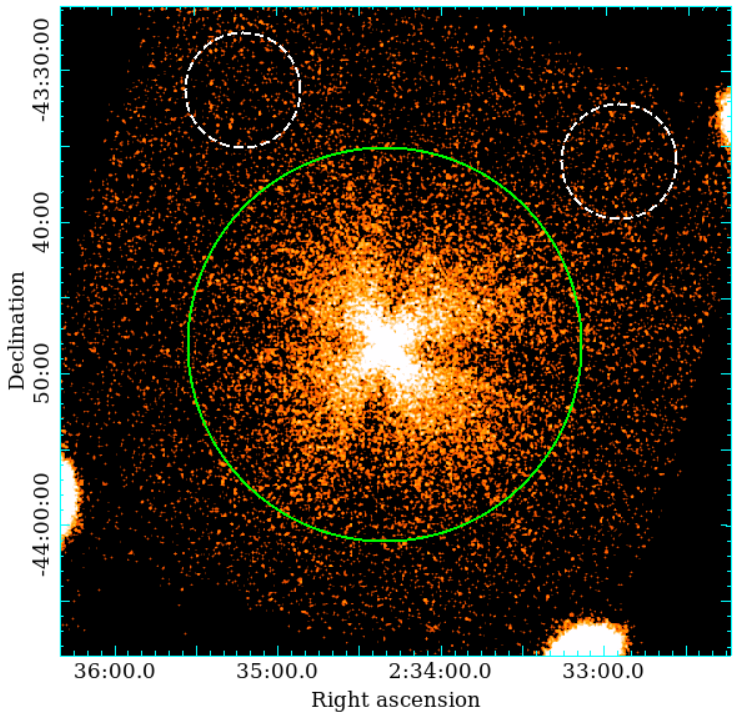}
\includegraphics[height=6.2cm, angle=-0]{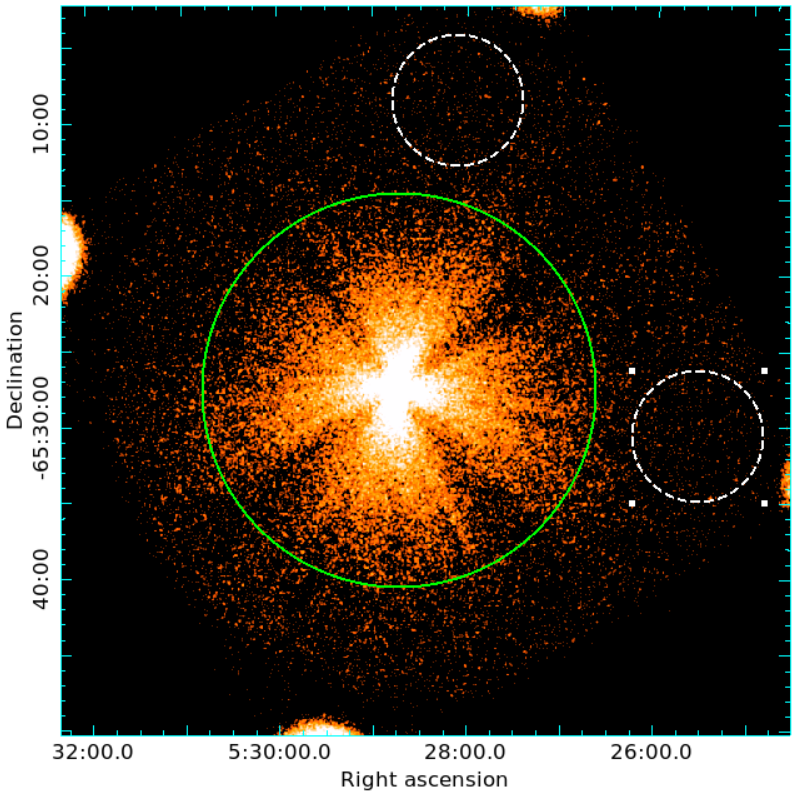}
\includegraphics[height=6.3cm, angle=-0, trim={0 0 0 3.3mm}]{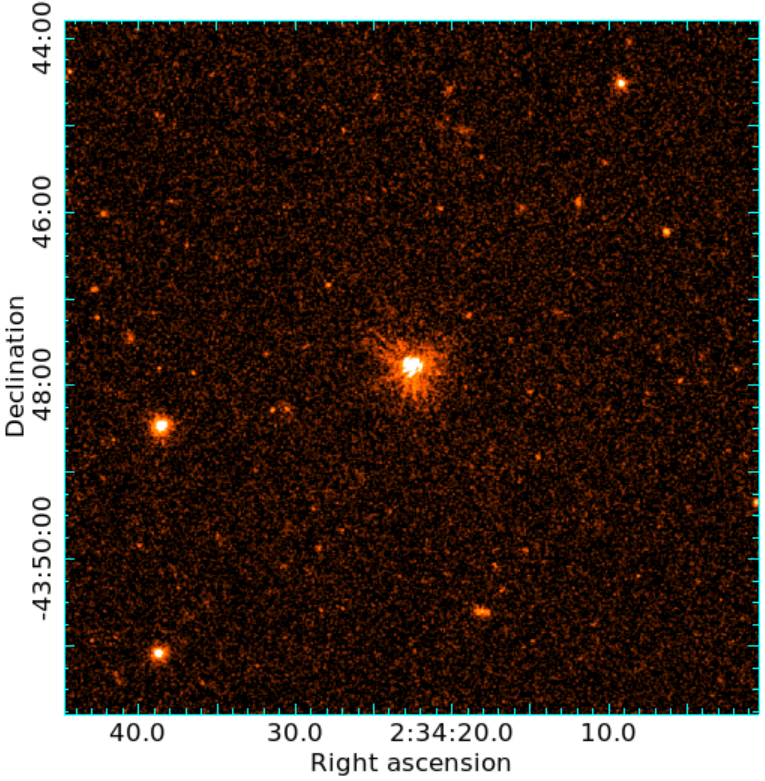}
\includegraphics[height=6.35cm, angle=-0]{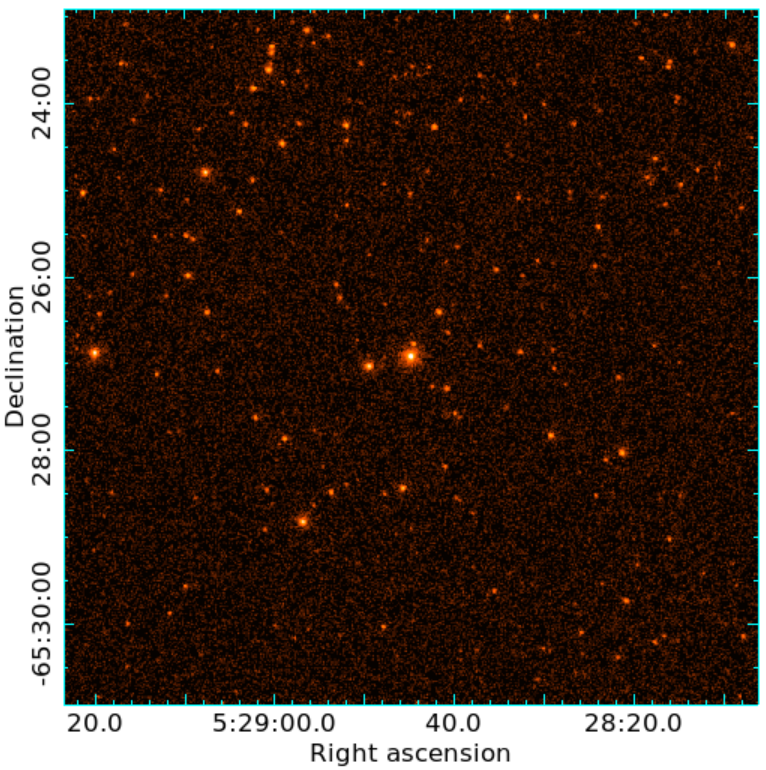}
\bigskip
\begin{minipage}{12cm}
\caption{\asts\ X-ray and UV images. In the top panel, the soft X-ray Images of \cce\ (left) and \abd\ (right) are shown. In the bottom panel, the  
near-ultraviolet (NUV) images of \cce\ (left) and far-ultraviolet (FUV) images of \abd\ (right) are shown with the sources placed at the center.
The selected source and background regions in the top panels are shown with green circles and white dashed circles, respectively.
}
\label{fig:image}
\end{minipage}
\end{figure*}

\section{Observations and Data Reduction}
\label{sec:redn}
We observed \cce\ and \abd\ using Soft X-ray focusing Telescope \citep[SXT;][]{SinghK-14-SPIE-1, SinghK-17-JApA-7}, Large Area X-ray Proportional Counter \citep[LAXPC;][]{AntiaH-17-ApJS}, Cadmium Zinc Telluride Imager \citep[CZTI;][]{BhaleraoV-17-JApA-3}, and Ultra-Violet Imaging Telescope \citep[UVIT;][]{TandonS-17-AJ-1}  onboard \asts\ observatory, on 2016 November 22 (PI. Karmakar; ID: A02\_151T01\_9000000818) and 2018 February 15 (PI. Karmakar; ID: A04\_116T01\_9000001896), respectively. In this article, we only report the results obtained by the SXT and UVIT instruments.  

\begin{figure*}
\centering
 \includegraphics[height=14.5cm, angle=-90, trim={7.1cm 0 0 0}, clip]{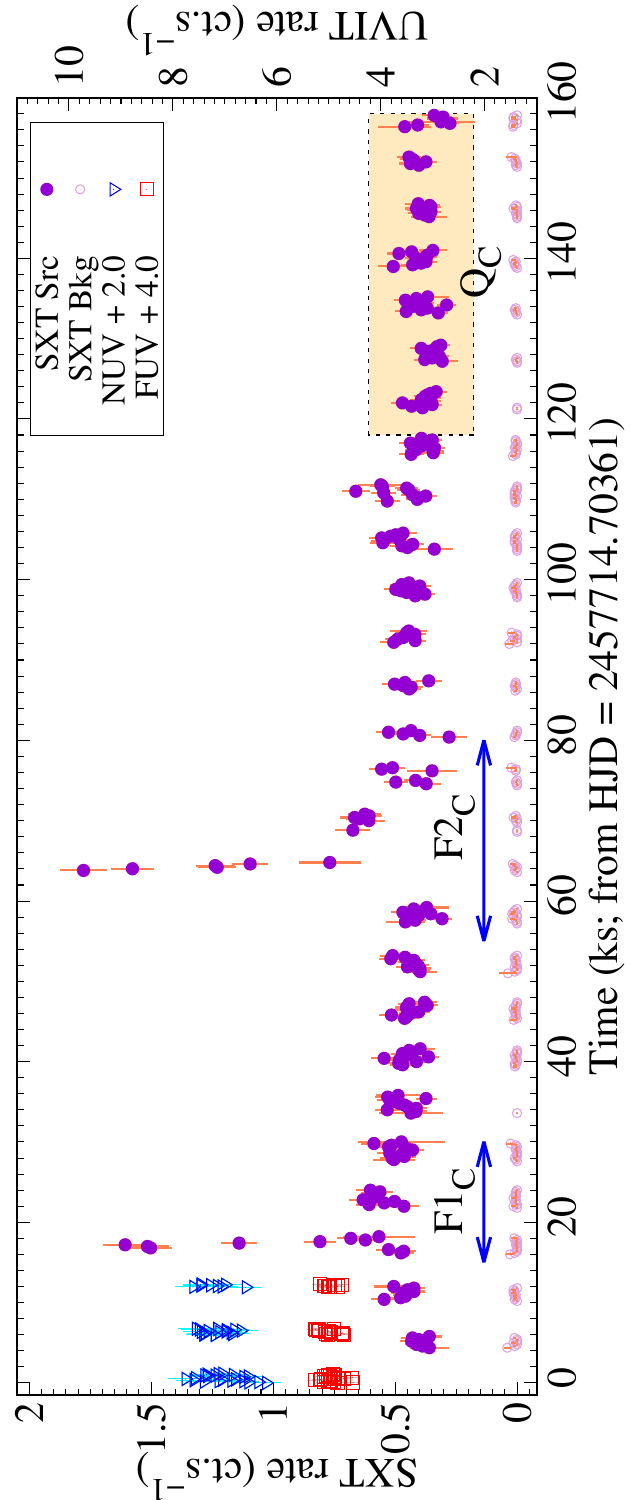}
 \includegraphics[height=14.5cm, angle=-90, trim={7.1cm 0 0 0}, clip]{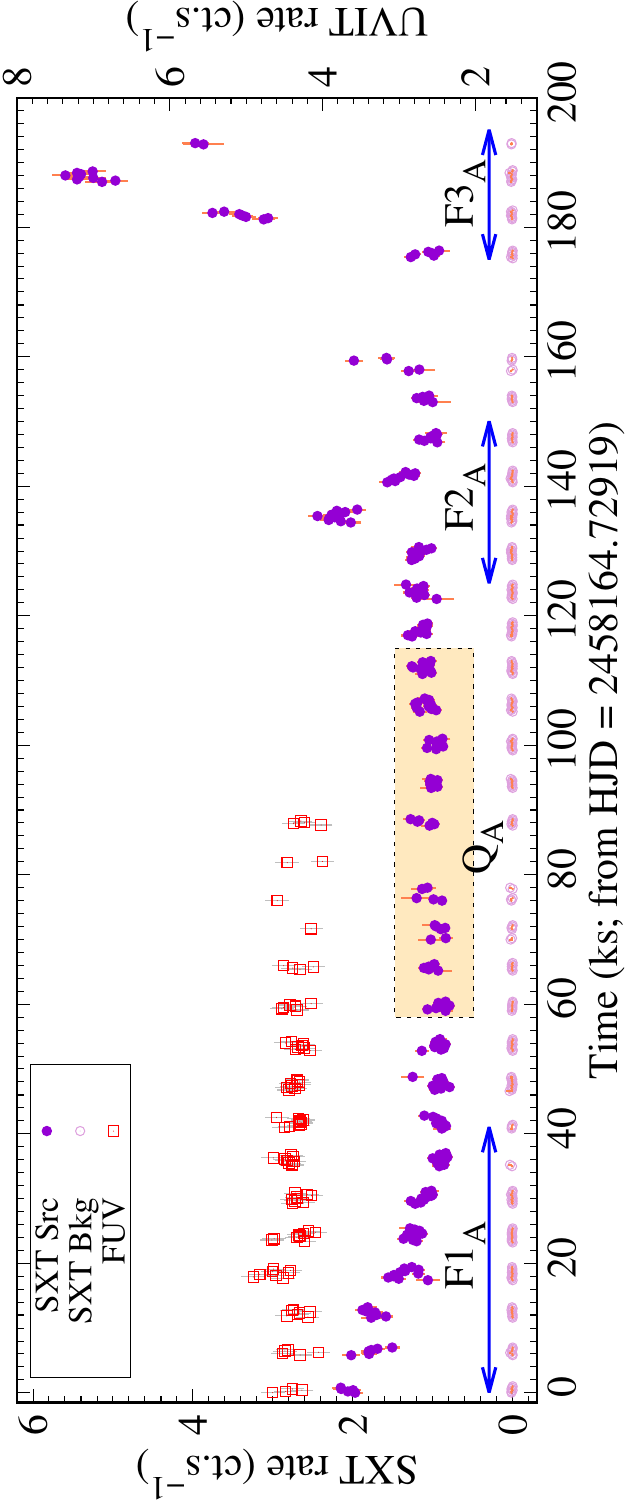}
\bigskip
\begin{minipage}{12cm}
\caption{X-ray and UV light curves from \asts\ SXT (left Y-axis) and UVIT (right Y-axis) have been shown. The top and bottom panel shows the background subtracted light curves of \cce\ and \abd, respectively, along with the background light curves for each observation. The arrows show the identified flare duration, whereas the yellow-shaded regions indicate the quiescent corona.}
\end{minipage}
\label{fig:lc}
\end{figure*}

\cce\ and \abd\ were observed in an energy range of 0.3--7 keV with SXT for a stare time of $\sim$160 and $\sim$200~ks. The level~1 data were processed using the \href{https://www.tifr.res.in/~astrosat_sxt/sxtpipeline.html}{\tt AS1SXTLevel2-1.4b} pipeline software to produce level~2 clean event files for each orbit of observation.
Each event file corresponding to each orbit was merged to a single cleaned event file using the \href{https://www.tifr.res.in/~astrosat_sxt/dataanalysis.html}{\tt sxtevtmerger} tool. In order to extract images, light curves, and spectra, we used the \href{https://heasarc.gsfc.nasa.gov/docs/software/ftools/ftools_menu.html}{\tt FTOOLS} task \href{https://heasarc.gsfc.nasa.gov/docs/software/lheasoft/ftools/xselect/xselect.html}{\tt xselect} V3.5, which has been provided as a part of the \href{https://heasarc.gsfc.nasa.gov/docs/software/heasoft/}{\tt heasoft} version 6.30. 
  Due to the broad PSF of \asts\ SXT (2\arcm\ inner and 10\arcm--12\arcm\ outer King's profile), we have chosen a circular region of a radius of 13\arcm\ centered at the source position to extract source products. We carefully verified that the region $\approx$17\arcm--19\arcm\ from the sources are not affected by the source brightness and can safely be considered as background. We, therefore, have chosen multiple circular regions of 2.$\!$\arcm6 radii as background regions and shown in Figure~\ref{fig:image}. The background light curves, as shown in Figure~\ref{fig:lc}, do not indicate any variation with the source. A similar background selection method was also followed by \cite{KarmakarS-22-MNRAS-1}.
The spectral analysis was carried out in an energy range of 0.3--7 keV using the X-ray spectral fitting package \citep[\href{https://heasarc.gsfc.nasa.gov/xanadu/xspec/}{\tt xspec};  version 12.11.1][]{Arnaud-96-ASPC-2}. For analysis of UVIT data and to extract the FUV and NUV light curve, we used the open-source \href{https://pypi.org/project/curvit/}{\tt curvit} python package \citep[][]{JosephP-21-JApA-1}. 

\begin{figure*}
\centering
\includegraphics[height=7.95cm, angle=-90, trim={0     0.1cm 0 0}, clip]{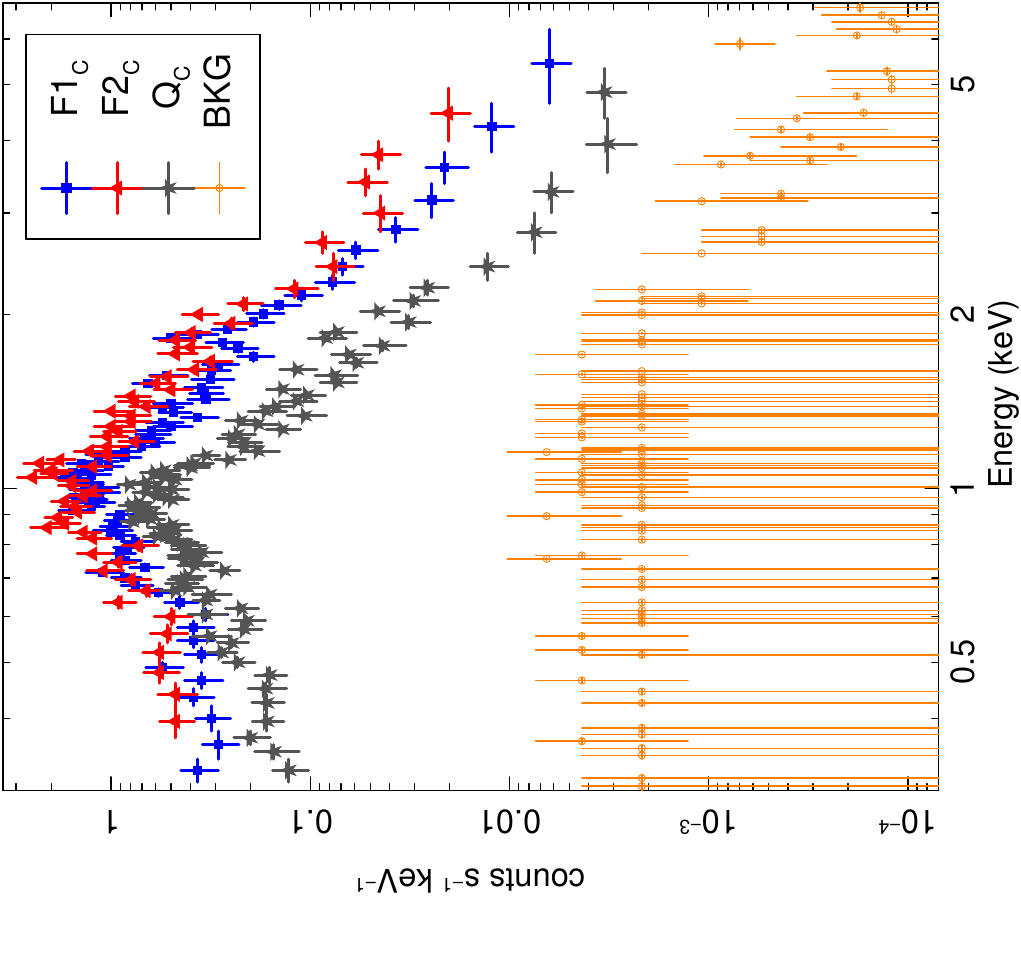}
\includegraphics[height=7.95cm, angle=-90, trim={0.5cm 0.1cm 0 0}]{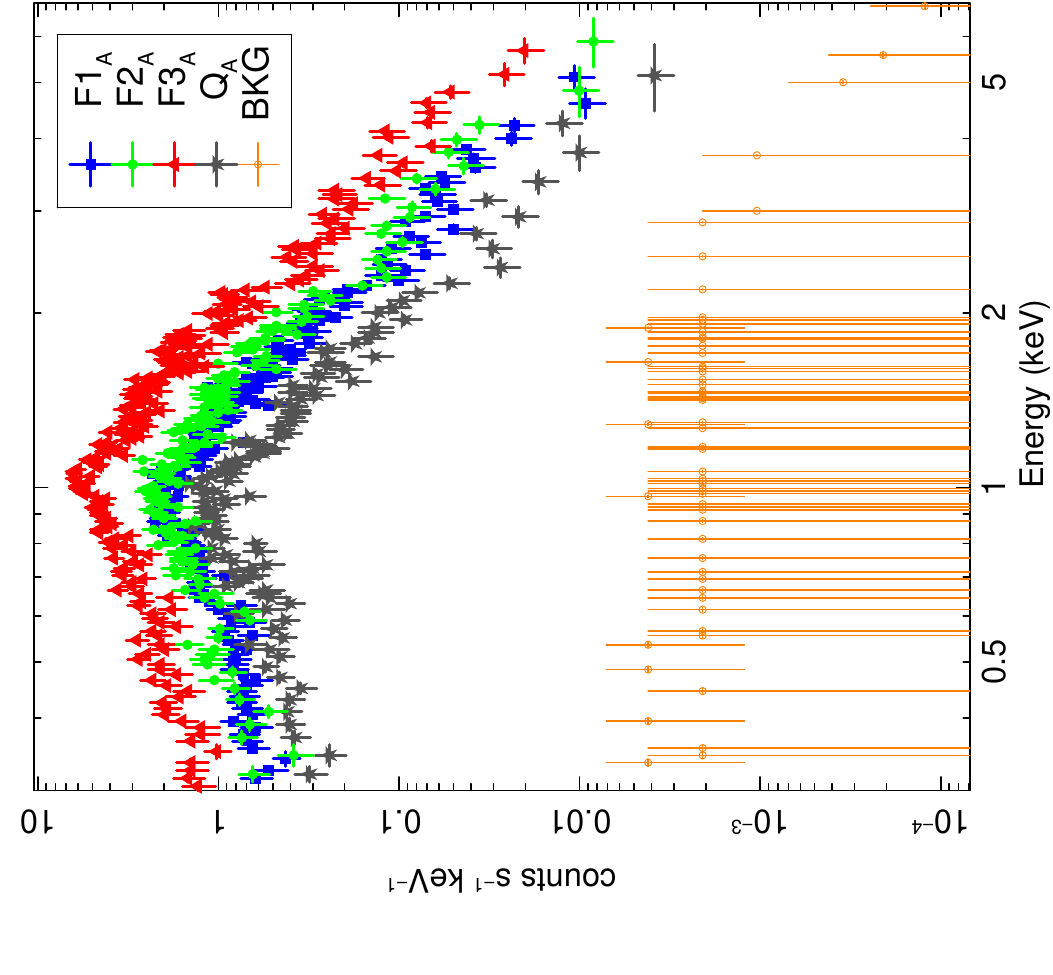}
\bigskip
\begin{minipage}{12cm}
\caption{The time-resolved \asts/SXT spectra for \cce\ (left) and \abd\ (right), along with the background spectra are shown. Different colors indicate the spectra correspond to different time segments, as shown in Figure~\ref{fig:lc}, whereas the background spectra are obtained from the complete observations.}
\label{fig:spectra}
\end{minipage}
\end{figure*}

\section{Analysis and Results}
\label{sec:result}
The background subtracted 0.3--7 keV X-ray light curves of \cce\ (top panel) and \abd\ (bottom panel) along with the background light curves are shown in Figure~\ref{fig:lc}.
The source light curves show a large variation in SXT count rates from 0.4 to 1.8 \cts\ for \cce, and 0.8 to 5.8 \cts\ for \abd. From SXT light curves we detected two flaring events for \cce\ (F1$_{\rm C}$ and F2$_{\rm C}$) and three flaring events are identified for \abd\ (F1$_{\rm A}$, F2$_{\rm A}$, and F3$_{\rm A}$). In Figure~\ref{fig:lc}, the UV light curves are also shown in the right-hand Y-axis. For \cce, both FUV and NUV observations were available, whereas \abd\ has been observed only in FUV. Unfortunately, both UV observation does not cover the whole observation time. For \cce, the observation was interrupted by \textit{Bright Object Dection} (BOD) phenomena at the onset of flare F1$_{\rm C}$, whereas the interruption in UV observation for \abd\ might be due to some technical difficulty. The FUV and NUV light curve for \cce\ and FUV light curve for \abd\ are found to remain constant within 1$\sigma$ uncertainty level.

We performed time-resolved spectroscopy for both sources with 11 segments for \cce\ and 25 segments for \abd. In Figure~\ref{fig:spectra}, we have shown the spectra of the quiescent state, the peak of each flare, and the background region for both observations. The background spectra show a similar pattern and background fluxes are multiple-order fainter than the quiescent spectra of the individual stars.
The stellar X-ray spectra have been analyzed with \textit{Astrophysical Plasma Emission Code} \citep[APEC; ][]{Smith-01-ApJ-96}, assuming the bremsstrahlung continuum and adopting the emission lines from the latest \textit{Astrophysical Emission Database}. Preliminary spectral analysis indicates the presence of three and four-temperature corona for CC Eri and AB Dor, respectively, where the highest temperature is found to vary with flare. The flare temperatures peaked at 51--59 MK for \cce\ and 29--44 MK for \abd. The peak emission measures of the flaring loops are estimated to be $\sim$10$^{54}$ for \cce\ and $\sim$10$^{55}$~cm$^{-3}$ for \abd. Global metallic abundances were found to vary with flare. The peak X-ray luminosities of the flares are found to be within 10$^{31-33}$~erg~s$^{-1}$.


\section{Discussion and Conclusion}
\label{sec:concl}
In this paper, we have investigated two active stars with the \asts\ observatory. A total of five flaring events have been detected. Although the stellar flares are spatially resolved, it is possible to infer the physical size and structure of the flares using the loop models. Assuming a semi-circular constant cross-section loop, the coronal loop height, plasma density, and the loop aspect ratio can be estimated using the quasi-static loop modeling \citep[][]{Van-den-oord-89-A+A-2}. Further investigation on the coronal loop properties and associated dynamo mechanisms is being carried out and will be presented elsewhere.

\begin{acknowledgments}
This publication uses data from the \asts\ mission of ISRO, archived at the Indian Space Science Data Centre (ISSDC). We thank the SXT Payload Operation Center at TIFR, Mumbai, for providing the necessary software tools. 
 This research has used the software provided by the High Energy Astrophysics Science Archive Research Center (HEASARC), which is a service of the Astrophysics Science Division at NASA/GSFC.
\end{acknowledgments}

\begin{furtherinformation}

\begin{orcids}
\orcid{0000-0001-8620-4511}{Subhajeet}{Karmakar}
\orcid{0000-0002-4331-1867}{Jeewan C.}{Pandey}
\orcid{0000-0002-4633-6832}{Nikita}{Rawat}
\orcid{0009-0002-6580-3931}{Gurpreet}{Singh}
\orcid{0000-0002-2489-5908}{Riddhi}{Shedge}
\end{orcids}

\begin{authorcontributions}
All authors have significantly contributed to this paper.
\end{authorcontributions}

\begin{conflictsofinterest}
The authors declare no conflict of interest.
\end{conflictsofinterest}

\end{furtherinformation}

\bibliographystyle{bullsrsl-en}

\bibliography{SK_collections}

\end{document}